\documentstyle[sprocl,epsfig]{article}

\def\PLB{{\em Phys. Lett.}  B}
\def\PRL{\em Phys. Rev. Lett.}
\def\PRD{{\em Phys. Rev.} D}

\def\be{\begin{equation}}
\def\ee{\end{equation}}
\def\bea{\begin{eqnarray}}
\def\eea{\end{eqnarray}}

\begin{document}

\title{
LEPTON FLAVOR VIOLATION IN THE HIGGS BOSON
DECAY AT A LINEAR COLLIDER\\}

\author{ 
SHINYA KANEMURA$^{1}$,
KOICHI MATSUDA$^{1}$,
TOSHIHIKO OTA$^{1}$,\\
TETSUO SHINDOU$^{2}$,
EIICHI TAKASUGI$^{1}$,
KOJI TSUMURA$^{1}$
 }

\address{
$^{1}$
Department of Physics, Osaka University, \\Toyonaka, Osaka
 560-0043, Japan\\
$^{2}$
Theory Group, KEK, Tsukuba, Ibaraki 305-0801, Japan
}


\maketitle\abstracts{
\hspace*{0.2cm} We study possibility of observing the process 
$h^{0} \rightarrow \tau^{\pm} \mu^{\mp}$
at a linear collider.
The branching ratio is constrained to be of the order of $10^{-4}$
by the $\tau^{-} \rightarrow \mu^{-} \eta$ result. 
Supersymmetric standard models can reproduce such amount of 
the branching ratio
by taking a specific parameter set.
The Higgsstrahlung process 
$e^{+}e^{-} \rightarrow Zh^{0}$
is preferable because of its simple kinematic
structure, 
then,
the signal process is $e^{+}  e^{-} \rightarrow Z h^{0} \rightarrow Z
\tau^{\pm} \mu^{\mp}$.
The most serious background comes from the process,  
$e^{+}  e^{-} \rightarrow Zh^{0} \rightarrow Z\tau^{\pm} \tau^{\mp}
\rightarrow \tau^{\pm} \mu^{\mp} \nu \bar{\nu}$.
We estimate the significance of the signal,
taking into account the background reduction.
}

\section{Introduction}
\hspace*{0.5cm}The search for the Lepton Flavor Violation (LFV) process is a promising way
to find the signal of new physics.
We propose a method to search for the LFV process
$h^{0} \rightarrow \tau^{\pm} \mu^{\mp}$
at a Linear Collider (LC).
We can measure the LFV Yukawa coupling directly by this process,
differently from the photon associated process $\tau\rightarrow\mu\gamma$ 
and the decays of a tau lepton $\tau \rightarrow\mu\mu\mu$ and $\tau
\rightarrow \mu\eta$.  

First, we show the experimental bound on the LFV Yukawa coupling.
The strongest bound comes from the result of the $\tau^{-} \rightarrow
\mu^{-}\eta$ search. 
Its upper limit can be realized in the Minimal
Supersymmetric Standard Model (MSSM)
with a specific parameter set.
Next, we consider the signal process $e^{+} e^{-} \rightarrow Zh^{0}
\rightarrow Z \tau^{\pm} \mu^{\mp}$ and also study the background
reduction.
There is an irreducible background which we call {\it the fake signal}.
We estimate the significance of the signal
taking into account the event numbers of the signal and the fake.
The details are shown in our recent work\,\cite{LFV-Higgs-LC}.

\section{LFV Yukawa Coupling}
\begin{figure}[th] 
     \begin{center}
     \epsfig{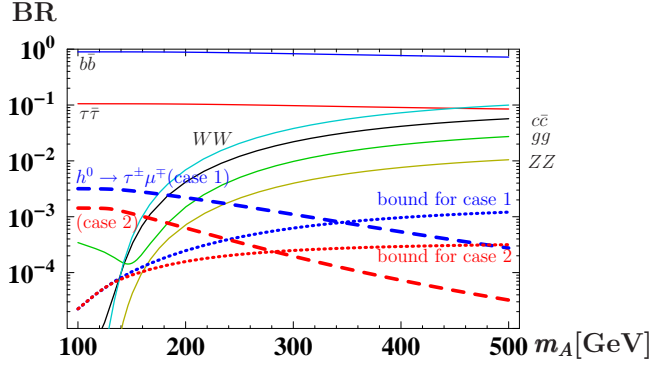}
     \end{center}
    \caption{The branching ratio of the $h^{0} \rightarrow \tau^{\pm} \mu^{\mp}$ 
    event for $m_{h}^{}=123$ GeV (dashed curve).
    The upper limit from $\tau^{-} \rightarrow \mu^{-}\eta$ result 
    is also shown (dotted curve).} 
     \label{Fig:Br-vs-mA}
   \end{figure}

\hspace*{0.5cm}The LFV Yukawa couplings are induced at the one-loop level in the MSSM.
The effective Lagrangian is written as follows\,\cite{Babu,Ellis,Rossi}:
\begin{eqnarray} 
 {\cal L}_{\tau \mu} = - \frac{ \kappa_{32} m_\tau}{v \cos^2\beta}
                 (\overline{\tau_{R}} \mu_{L}) 
 \left\{ \cos(\alpha-\beta) h^0 + \sin(\alpha-\beta) H^0 - {\rm i} A^0 
\right\} + {\rm H.c.}. 
\label{eq:effLagrangian}
\end{eqnarray}
Consequently, the branching ratio for $h^{0} \rightarrow \tau^{\pm} \mu^{\mp}$ is 
approximately given by\,\footnote{A full 1-loop calculation is shown in Ref.[5].}  
\begin{eqnarray} 
 {\rm Br}(h^{0} \to \tau^{\pm}\mu^{\mp}) \sim \frac{1}{N_c}  \frac{m_\tau^2}{m_b^2} 
        \frac{\cos^2(\alpha-\beta)}{\cos^2\beta\sin^2\alpha} 
        \times |\kappa_{32}|^2. \label{eq:Br-h-TauMu}
\end{eqnarray} 
The LFV parameter for this interaction, $|\kappa_{32}|$, is experimentally 
constrained by the result of  
$\tau^{-}\rightarrow \mu^{-}\eta$ as\,\cite{Sher,Belle}, 
\begin{eqnarray}
 |\kappa_{32}|^2  <   
  0.3 \times 10^{-6} \times 
  \left(\frac{m_{A}^{}}{150 {\rm GeV}}\right)^4
  \left(\frac{60}{\tan \beta}\right)^{6}.
 \label{lim}
\end{eqnarray}
This shows that the bound is relaxed for larger $m_A^{}$ and smaller $\tan\beta$. 
The LFV parameter $|\kappa_{32}|$ is a function of the SUSY parameters, and 
its value does not depend on the
absolute values of the SUSY parameters but on their ratio $\mu /
m^{}_{\rm SUSY}$, where $\mu$ is the higgsino mass and 
$m^{}_{\rm SUSY}$ is the typical scale of the SUSY
particles\,\cite{Rossi}.
Therefore, 
the effect of this interaction does not decouple
even in the large $m^{}_{\rm SUSY}$ limit
as long as the ratio is set to be of $\mathcal{O}$(1).
Let us consider the following parameter sets: \vspace{0.1cm}\\
\hspace{0.3cm}
\begin{minipage}{12cm}
\begin{itemize}
\item[Case 1:]
$\tan\beta=60$, $\mu=25$ TeV, \vspace{-0.1cm}\\
$M_1 \sim M_2 \sim m_{\tilde{\ell}_{{L}_{\mu,\tau}^{}}}^{} \sim 
 m_{\tilde{\ell}_{{R}_{\mu,\tau}^{}}}^{} \sim
 m_{\tilde{\nu}_{L_{\mu,\tau}^{}}}^{} \sim  
 \sqrt{\bigl|(\Delta m_{\tilde{l}_L^{}}^2 )_{32}^{}\bigr|} \sim 2$ 
TeV, \\  
$M_{Q}^{} \sim 10$ TeV and $M_{U,D}^{} \sim A_{t,b} \sim 8$ 
TeV, 
\item[Case 2:]
$\tan\beta=60$,  $\mu=10$ TeV,  \\
$m_{\tilde{\ell}_{{L}_{\mu,\tau}^{}}}^{} \sim
 m_{\tilde{\nu}_{L_{\mu,\tau}^{}}}^{} \sim 
\sqrt{\bigl|(\Delta m_{\tilde{l}_L^{}}^2 )_{32}^{}\bigr|} \sim 1.2$ TeV, 
$m_{\tilde{\ell}_{{R}_{\mu,\tau}^{}}}^{} \sim 0.9$ TeV,\\ 
$M_1 \sim 1$ TeV, $M_2 \sim 0.8$ TeV, 
$M_{Q}^{} \sim 5$ TeV and $M_{U,D}^{} \sim A_{t,b} \sim 3$ 
TeV,
\end{itemize}
\end{minipage}\vspace{0.1cm}\\
where
$M_{1,2}$ are the gaugino masses for $U(1)_{Y}$ and $SU(2)_{L}$, respectively,
$m_{\tilde{\ell}}$ and $m_{\tilde{\nu}}$ are the masses of the charged
slepton and the sneutrino, $M_{Q}$, $M_{U}$, $M_{D}$, and $A_{t,b}$ are the
soft SUSY breaking parameters 
for the squark sector. 
In these examples,
the photon-associated penguin diagrams decouple, and only the
Higgs-mediated LFV coupling can contribute to 
$\tau^{-} \rightarrow \mu^{-}\eta$.
 For Case 1 and Case 2, we obtain 
$|\kappa_{32}|^{2} \sim 8.4 \times 10^{-6}$ and $3.8 \times 10^{-6}$,
and they are allowed experimentally for $m_{A}^{} > 350$ GeV and $m_{A}^{}>280$
GeV, respectively.
The branching ratio can be as large as $10^{-4}$ 
(See Fig.\ref{Fig:Br-vs-mA}).
The mass of the lightest Higgs boson is fixed to be 123 GeV in both cases.

\section{Signal at a Linear Collider}
     \begin{figure}[th] 
     \begin{center}
      \epsfig{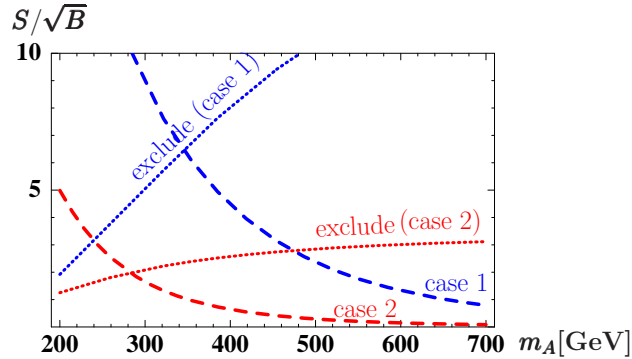}
     \end{center}
     \caption{The significance of the signal event (dashed curve).
      The upper limit from the $\tau^{-} \rightarrow \mu^{-} \eta$ result is
      also shown (dotted curve).} 
     \label{Fig:significance}
     \end{figure}
\hspace*{0.5cm}In order to study the LFV Higgs decay, 
the Higgsstrahlung process is appropriate 
because of its simple kinematic structure:
\begin{equation}
e^{+} e^{-} \rightarrow Z h^{0} \rightarrow Z \tau^{\pm} \mu^{\mp}.
\end{equation}
We can identify the signal by using the recoil momentum of the $Z$ boson,
without measuring the momentum of the tau lepton.
A large part of the background can be reduced by the invariant mass cut. 
However, there remain irreducible backgrounds which we call {\it
the fake signals}.
They are induced via
\begin{equation}
e^{+}  e^{-} \rightarrow Z h^{0} \rightarrow Z \tau^{\pm} \tau^{\mp} 
 \rightarrow  Z \tau^{\pm} \mu^{\mp} \nu \bar{\nu}.
\label{eq:background}
\end{equation}
Among the events of Eq.(\ref{eq:background}), 
those which mimic the $\tau$-$\mu$ pair
production are regarded as the fake signals.
We assume that the integrated luminosity is 1,000 fb$^{-1}$ 
and that $\sqrt{s}$ is optimally tuned for the Higgs production via 
the Higgsstrahlung.
We categorize the signal $Z\tau\mu$ into two groups depending on the decay
products of the $Z$ boson: 
one is $jj\tau\mu$ where $j$ is the hadronic jet, 
and the other is $\ell\ell\tau\mu$ where $\ell$ denotes electron or muon.
The resolution of the $Z$ boson momentum reconstructed from the hadronic jets is
expected to be 3 GeV and that from the lepton pair is 1 GeV, respectively.  
In Case 1, the numbers of $jj\tau\mu$ and $\ell\ell\tau\mu$ are evaluated as 
$N^{\rm signal}_{jj\tau\mu}=118$ and $N^{\rm signal}_{\ell\ell\tau\mu}=11$.
The fake signal is also evaluated as $N^{\rm fake}_{jj\tau\mu}=460$ and 
$N^{\rm fake}_{\ell\ell\tau\mu}=15$.
The combined significance, $N^{\rm signal}/\sqrt{N^{\rm fake}}$, 
can reach to 6.3 at $m_{A}^{}\simeq350$ GeV.
In Case 2, the significance is at most large as 2.0 at $m_{A}\simeq280$ GeV.
    
\section{Summary}
\hspace*{0.5cm}We have studied possibility for the direct measurement of the LFV
Yukawa coupling at a LC in the framework of the MSSM.
The direct measurement of the LFV Yukawa coupling 
could be complementary to the measurement of the other types of 
the LFV couplings.
We have estimated the significance of the signal process,
$e^{+}  e^{-} \rightarrow Zh^{0} \rightarrow Z\tau^{\pm} \mu^{\mp}$, 
at a LC with a high luminosity.
The signal can be identified
by using the recoil of the $Z$ boson
without measuring the momentum of the tau lepton.
A large part of the backgrounds is expected to be reduced by appropriate
kinematic cuts.
We have found that the signal can be marginally visible. 
Needless to say, a more realistic simulation analysis is necessary.

{\it Acknowledgements}:
This work was supported, in part, by JSPS Research Fellowship 
for Young Scientists (No.15-3693, 15-3700, 15-3927). 
%



\section*{References}
\begin{thebibliography}{99}
\bibitem{LFV-Higgs-LC}
	S. Kanemura, K. Matsuda, T. Ota, T. Shindou, E. Takasugi,
	K. Tsumura,
	arXiv:hep-ph/0406316, to appear in \PLB.

\bibitem{Babu}  K.S.~Babu, and C.~Kolda, \PRL \rm {\bf 89} (2002) 241802. 

\bibitem{Ellis} A.~Dedes, J.~Ellis, and M.~Raidal, 
                 \PLB \rm {\bf 549} (2002) 159. 

\bibitem{Rossi}
	A.~Brignole, and A.~Rossi,
         \PLB \rm {\bf 566} (2003) 217.
        A.~Brignole, A.~Rossi,
        arXiv:hep-ph/0404211.

\bibitem{Herrero}
	E.~Arganda, A.~M.~Curiel, M.~J.~Herrero and D.~Temes,
	arXiv:hep-ph/0407302.


\bibitem{Sher}
        M.~Sher, 
        \PRD \rm {\bf 66} (2002) 057301.

\bibitem{Belle}
	Y. Enari, {\it et al.},
	the Belle collaboration
	arXiv:hep-ph/0404018.

\end{thebibliography}
\end{document}